\documentclass[letterpaper]{article} 
\usepackage{aaai25}  
\usepackage{times}  
\usepackage{helvet}  
\usepackage{courier}  
\usepackage[hyphens]{url}  
\usepackage{graphicx} 
\usepackage{natbib}  
\usepackage{caption} 
\usepackage{algorithm}
\usepackage{algorithmic}
\usepackage{subcaption}
\usepackage{amsmath,amssymb}
\usepackage{tikz}
\usepackage{quantikz}

\usepackage{newfloat}
\usepackage{listings}
\DeclareCaptionStyle{ruled}{labelfont=normalfont,labelsep=colon,strut=off} 
\floatstyle{ruled}
\newfloat{listing}{tb}{lst}{}
\floatname{listing}{Listing}

\nocopyright

\title{Quality Diversity for Variational Quantum Circuit Optimization \footnote{Preprint of work accepted to appear in Proceedings of the 35th International Conference on Automated Planning and Scheduling (ICAPS 2025)}}
\author{
    Maximilian Zorn\textsuperscript{\rm 1},
    Jonas Stein\textsuperscript{\rm 1,2}, 
    Maximilian Balthasar Mansky\textsuperscript{\rm 1},\\ 
    Philipp Altmann\textsuperscript{\rm 1}, 
    Michael Kölle\textsuperscript{\rm 1}, 
    Claudia Linnhoff-Popien\textsuperscript{\rm 1}
}
\affiliations{
    \textsuperscript{\rm 1}LMU Munich\\
    \textsuperscript{\rm 2}Aqarios GmbH\\
}

\begin{document}

\maketitle

\begin{abstract}
    Optimizing the architecture of variational quantum circuits (VQCs) is crucial for advancing  quantum computing (QC) towards practical applications. Current methods range from static ansatz design and evolutionary methods to machine learned VQC optimization, but are either slow, sample inefficient or require infeasible circuit depth to realize advantages. Quality diversity (QD) search methods combine diversity-driven optimization with user-specified features  that offer insight into the optimization quality of circuit  solution candidates. However, the choice of quality measures and the representational modeling of the circuits to allow for optimization with the current state-of-the-art QD methods like covariance matrix adaptation (CMA), is currently still an open problem. In this work we introduce a directly matrix-based circuit engineering, that can be readily optimized with QD-CMA methods and evaluate heuristic circuit quality properties like expressivity and gate-diversity as quality measures. We empirically show superior circuit optimization of our QD optimization w.r.t. speed and solution score against a set of robust benchmark algorithms from the literature on a selection of NP-hard combinatorial optimization problems.
\end{abstract}

\section{Introduction}\label{sec:introduction}

As the field of quantum computing continues to grow -- even with the limitations of recent noisy intermediate scale quantum (NISQ) hardware \citep{preskill2018quantum} -- ideas for utilizing the potential computing speedup of quantum computing have been tested for a wide range of problems \citep{ronnow2014defining}. Some examples include application in many different fields of study, from financial prediction ~\citep{egger2020finance}, fraud detection \citep{kyriienko2022unsupervised}, image classification \citep{senokosov2023quantum} and quantum machine learning (QML) (cf. \citep{biamonte2017quantum}), material science \citep{kandala2017hardware}, chemistry simulations \citep{cao2019quantum} and combinatorial optimization \citep{khairy2020learning,li2020quantum}.

In this paper we focus on combinatorial optimization (CO), as a branch of mathematical optimization that seeks to find the best solution from a finite set of discrete possibilities. CO plays a critical role in various fields, including computer science, operations research, and engineering, by addressing complex problems such as scheduling, network design, and resource allocation. Despite its wide applicability, combinatorial optimization often faces significant computational challenges, necessitating the development of efficient algorithms and heuristics to achieve optimal or near-optimal solutions.

However, leveraging the potential capabilities of quantum hardware promises for CO may eventually lead to tangible speedups, as long as quantum circuits enabling this speedup can be constructed. Current methods for designung such circuits range from analytic ansatz design to dynamically constructed variational quantum circuits (VQCs) \citep{cerezo2021variational}. VQCs are characterized by their use of parameterized rotation gates, the positioning, and (parameter) tuning of which is usually done via search-based optimization methods. VQC optimization can be mainly divided in gradient-based and gradient-free methods. While the former is effective for small circuits, the phenomenon of \textit{barren plateaus} \citep{larocca2024review}, where the optimization landscape of gradient-based optimizers becomes mostly flat and featureless and therefore hard to adjust, makes the optimization of deeper circuits a challenge, in particular in combination with the current noisy hardware \citep{wang2021noise}. Gradient-free methods, on the other hand, can avoid this optimization pitfall of barren-plateaus through diverse populations of individuals or evolutionary methods (e.g., \citet{sunkel2023ga4qco, giovagnoli2023qneat}), but are often slow and sample inefficient, which makes testing both time- and cost-inefficient.

In this work, we enable the integration of quality diversity (QD) driven optimization to the problem of optimizing VQCs. In essence, QD is a universally applicable search method that explores diverse solutions for a given objective function based on low dimensional user-defined characteristic (quality) measures. While QD has many recent examples of successful utilization \citep{fioravanzo2019constrained, cully2015robots, zhang2024arbitrarily}, to the best of our knowledge, QD has not yet been developed for the optimization of VQCs. We suspect this is due to a lack of low dimensional selection of quality measures and the way circuits are represented, which are essential for optimization using current state-of-the-art quality diversity methods like Covariance Matrix Adaptation (CMA). As such, VQC QD-design has remained an unresolved problem, possibly due to the absence of a standardized circuit encoding that would facilitate more efficient circuit search.\\

In this paper we resolve both these issues to enable VQC optimization via quality-diversity CMA methods. We see the key contributions of this work in the following aspects:
\begin{itemize}
    \item We propose a lightweight but efficient $1:1$ encoding of whole quantum circuits into optimizable matrices, which we utilize for the optimization of VQCs on a set of multiple well-known combinatorial optimization problems. We leverage recent state-of-the-art QD methods into our search method, defining NISQ-friendly quality metrics for \textit{circuit expressivity} -- heuristically approximated by \textit{circuit sparsity} -- and \textit{gate diversity}. The encoding itself, additionally, is designed to be independent of specific quantum hardware, problem specific aspects or domain knowledge and is thus generally available for any kind of optimization.
    
    \item We demonstrate the capability of our approach by achieving superior results on combinatorial optimization problems -- both in terms of objective function quality and search speed -- in direct comparison to two kinds of state-of-the-art circuit construction approaches, a different gradient-free evolutionary VQC optimization method (QNEAT) and the analytically designed circuit approximation optimization algorithm (QAOA).
\end{itemize}

\section{Related work}\label{sec:related_work}
While there are works exploring quantum ML~\citep{biamonte2017quantum}, quantum neural networks~\citep{beer2020training}, and quantum reinforcement learning~\citep{dong2008quantum, dunjko2017advances} running entirely \textit{on} the quantum computer, the gate cost for setup and logic on current QC hardware is still the constraining factor for realizing their full potential. Instead, a more realistic approach to circuit construction and optimization -- for specific problems in traceable sizes -- currently happens in a hybrid fashion: For quantum gate-based computing \citep{nielsen2010quantum},  usually, a specific ansatz design is chosen or dynamically assembled, and the parameterization is then learned by a classical optimization method.

Reinforcement learning (RL), traditionally successful in domains requiring sequential decision-making under uncertainty, also provides promising ideas for automating and optimizing quantum circuit design. Hence, as a field of study, quantum reinforcement learning (QRL) \citep{dong2008quantum, meyer2022survey, jerbi2021parametrized} has recently gained traction for e.g., its application to quantum state preparation \citep{wu2020quantum, gabor2022applicability} or for solving classical RL with quantum agents \citep{skolik2022quantum, chen2020variational, lockwood2020reinforcement}. 

A related discipline lies in the machine-learned (ML) predicted parametrizations for established gate sequences or rotation blocks, like the variational ansatz (VA). An overview of different VAs can be found in \citep{cerezo2021variational} for instance. More recently, the work of \citep{tilly2022variational} also concisely covers VQC best practices, which inspiration can be taken from. Use cases include, e.g., solving linear equations~\citep{bravo2019variational} and modeling/training agents for advanced RL environments (frozen lake, cognitive radio)~\citep{chen2020variational} on quantum hardware. 

More unconstrained approaches with RL/ML for specific problems are also getting more common; \citep{an2019deep} used deep Q-learning networks to predict time-sensitive quantum gate control for single qubit Hadamard gates. \citep{ostaszewski2021reinforcement} used RL to learn an optimized composition of rotation gates by treating the quantum circuit like a grid with gates on qubit positions over time, which they then optimize to estimate the ground state energy of lithium hydride. \citep{fosel2021quantum} go in the opposite direction and opt for an approach of optimizing the arrangement of whole, randomly pre-sampled quantum circuits instead. In their case the agent learns to select discrete actions for changing or simplifying the arrangement of gates from a set of possible transformations. \citep{mackeprang2020reinforcement} utilize RL for quantum state engineering, where they focus learning maximally entangled 2-qubit states, using a discrete action agent (deep Q-network, DQN) and let it choose from seven predefined spin/projection options that modify the current circuit state. Similarly, the work of \citep{kolle2024reinforcement} proposes the use of RL as a gate classification task, where the agent chooses from a selection of gates to place at certain positions on the circuit. In contrast to these single-step-single-gate approaches, we instead opt to optimize the circuit in its entirety.

\section{Background}\label{sec:background}

\subsection{Quantum Circuits}
Current quantum computing (cf. \citet{nielsen2010quantum}) and their computational circuits rely on the successive application of gates to states. A state carries information, while the gates manipulate the information in correspondence to a quantum algorithm. Numerically, states are represented as complex-valued vectors of dimension $2^n$, $n$ the number of qubits. The quantum operation $U$ can be represented by a matrix of $2^n \times 2^n$ dimensions. Individual gates within the operation often have lower-dimensional representations that are concatenated with the tensor product.

Our work approaches this tensored product of gates. In the graphical picture of quantum circuits, our approach can be thought of as the horizontal concatenation of $L$ layers of $n$ vertical tensored gates, for a dense grid of operations. This grid structure is natural to conventional optimization and abstracts the quantum circuit construction.

Moving beyond a specific number of qubits, we generalize our model to a \( n \)-qubit setting. In this more general framework, instead of considering only a fixed number of amplitudes or states, we must account for the probability of measuring any of the \( 2^n \) possible state combinations in the quantum system. The state of an \( n \)-qubit quantum circuit can be expressed as: $|\theta\rangle = \sum_{i=0}^{2^n-1} \alpha_i |i\rangle$, where \( |i\rangle \) represents each possible basis state of the \( n \)-qubit system, encoded in binary notation (e.g., \( |0\rangle, |1\rangle, |2\rangle, \ldots, |2^n-1\rangle \)). Here, \( \alpha_i \) are the complex amplitudes corresponding to each state, and \( i \) ranges over all possible states. The vector representation of the corresponding Dirac notation for this state is then  $\theta\rangle = [\alpha_0, \alpha_1, \ldots, \alpha_{2^n-1}]^\intercal$.

Finally, to satisfy the fundamental rule of probability, the sum of the squares of the magnitudes of these amplitudes must equal 1, ensuring the state vector is properly normalized. This condition is formulated as 
$\sum_{i=0}^{2^n-1} |\alpha_i|^2 = 1$.

\subsection{Combinatorial Optimization}\label{subsec:maxcut}
In this paper we will evaluate our QD approach on the four following combinatorial binary optimization graph problems, all of which are frequently utilized as quantum optimization benchmarks, e.g., \citep{khairy2020learning}. To test for selection or inclusion from the set of vertices $V(G)$ or set of edges $E(G)$ (or its complement $E(\bar{G})$), we measure our solution candidate circuits with $Z_i \to [-1, 1]$, as it is commonly done, where $Z$ is the Pauli-Z operator acting on the qubit corresponding to the vertice $i$ of a graph $G$. The tensor product \( Z_i \otimes Z_j \) computes the interaction between two vertices connected by edge $(i,j)$.

\begin{itemize}
    \item The maximum cut (\textit{MaxCut}) problem devides nodes on a graph into two separate sets (with labels $-1$ or $1$) so that the splitting line goes through the maximum number of edges in the graph $G$. The cost hamiltonian to maximize is given as: $$\max - H_{MC} = \sum_{(i,j)\in E(G)} \frac{1}{2}(Z_i \otimes Z_j - 1)$$

    \item The minimum vertex cover (\textit{MinVEC}) problem involves identifying the smallest set of vertices in a graph such that every edge is incident to at least one vertex in the set, via cost hamiltonian: $$\max - H_{VC} = 3 \sum_{(i,j)\in E(G)} (Z_i \otimes Z_j + Z_i + Z_j) - \sum_{i \in V(G)} Z_i$$

    \item The maximum independent set (\textit{MaxIND}) problem seeks the largest subset of vertices in a graph where no two vertices are adjacent with: $$\max - H_{IS} = 3 \sum_{(i,j)\in E(G)} (Z_i \otimes Z_j - Z_i - Z_j) + \sum_{i \in V(G)} Z_i$$

    \item The maximum clique (\textit{MaxCLI}) problem aims to find the largest complete subgraph within the given graph, where every pair of vertices in the subgraph is directly connected by an edge. The cost hamiltonian penalizes the complement of all selected edges $E(\bar{G})$ with: $$\max - H_{CL} = 3 \sum_{(i,j)\in E(\bar{G})} (Z_i \otimes Z_j - Z_i - Z_j) + \sum_{i \in V(G)} Z_i$$
\end{itemize}

\subsection{Quality Diversity Methods}
Quality diversity methods are a family of gradient-free single-objective optimization algorithms. The diversity aspect of QD is inspired by evolutionary algorithms with diversity optimization such as novelty \cite{lehman2008exploiting, lehman2011abandoning, conti2018improving}, which similarly optimize an objective function and diversify a set of diversity measure functions to generate a diverse collection of high-quality solutions. 

\textbf{QD algorithms} have many different interpretations, incorporating different optimization methods, such as topological search (cf. NEAT \cite{stanley2002evolving}), gradient descent \cite{fontaine2021differentiable, fontaine2023covariance} or model-based surrogates \cite{gaier2018data}. Many of the more recent QD methods are based on the concept of \textit{MAP-Elites} \cite{mouret2015illuminating}, which moves away from pure novelty search in favor of exploring a high-dimensional space (of an objective function) with the intention of finding high-performing solutions at each point in a low-dimensional quality-measure space, where the user gets to choose quality dimensions of interest. The collection of diverse and qualitative solutions is referred to as an \textit{Archive}, containing the best performing combination of measure-representatives, called \textit{elites}, in each discretized cell. Additionally, the \textit{QD-score} measures the quality and diversity of the elites by summing the objective values of elites in the archive. After a fixed amount of exploration, the set of elites is returned. Application of MAP-Elites can be found in, e.g., constrained optimization \citep{fioravanzo2019constrained}, robotic adaptability \cite{cully2015robots} or arbitrarily scalable environment generators \cite{zhang2024arbitrarily}.

\textbf{Covariance Matrix Adaptation} From the literature of QD methods, we explore the methods of Covariance Matrix Adaptation (CMA) \citep{hansen2016cma} as our circuit optimization algorithm. CMA, initially based on evolutionary strategies (ES), maintains a population of solution samples (generation) and moves each iteration toward the center of the highest objective evaluation.

CMA, in its simplest form \textbf{CMA-ES}, models the sampling distribution of the population as a multivariate normal distribution $\mathcal{N}(m, C)$, where $m$ is the distribution mean and $C$ is its covariance matrix. The main mechanisms steering CMA-ES are the selection and ranking of the $\mu$ fittest solutions, which update the next generation's next sampling distribution. A history of aggregate changes to $m$ (the evolution path) provides information about the search process similar to momentum in stochastic gradient descent. To avoid the quick convergence of CMA-ES, the exploration aspects of the Map-Elites algorithm were integrated into the \textbf{CMA-ME} version \cite{fontaine2020covariance} to create a population of modified CMA-ES instances called \textit{emitters}, that each performs a search with feedback gained from interacting with the archive. The concept of emitters extends CMA-ES by adjusting the ranking rules that form the covariance matrix update to maximize the likelihood that future steps in a given direction result in archive improvements with respect to the quality measures.

Finally, CMA MAP-Annealing (\textbf{CMA-MAE}) \cite{fontaine2020covariance,fontaine2023covariance} has emerged with state-of-the-art performance in continuous domains. CMA-MAE extends MAP-Elites by incorporating the self-adaptation mechanisms of CMA-ES. CMA-ES maintains a Gaussian distribution, samples from it for new solutions, evaluates them, and then updates the distribution towards the high-objective region of the search space. Furthermore, an \textit{optimization archive} updating mechanism to balance exploitation and exploration of the measure space is maintained alongside the \textit{result archive}. The mechanism introduces a threshold value $t_e$ (independent of the objective measure) to each cell $e$ in the archive, which determines whether a new solution $\theta'$ should be added. New solution $\theta'$ is then accepted iff. $f(\theta') > t_e$, where $f(\theta')$ is the objective evaluation of $\theta'$. The threshold values are iteratively updated via an archive learning rate $\alpha \in [0;1]$ (to infer the `improvement rate' of the search at cell $e$), calculated as $\Delta = f(\theta') - t_e$. Upon acceptance in the respective cell, $t_e$ is updated via $t_e \leftarrow (1-\alpha)t_e + \alpha f(\theta')$, controlled by learning rate $\alpha$. 

In choosing $\alpha$, we can adjust the CMA framework on a `scale' from CMA-ES to CMA-MA: With $\alpha = 1$, CMA-MAE behaves like CMA-ME, with the improvement control greedily moving away from diminishing improving iterations. On the other hand, with $\alpha = 0$, CMA-MAE behaves like CMA-ES since improvement values always correspond to the objective values in the $t_e$ update term. As the threshold will never change, CMA-MAE will only optimize the objective, akin to the evolutionary strategy CMA-ES. With $\alpha$ values between $0$ and $1$, CMA-MAE will gradually anneal the exploration through $t_e$ and `linger' around promising solutions, even if their improvement rate diminishes.

\section{Method}\label{sec:method}

\subsection{Quantum Circuit Encoding}\label{subsec:circuit_encoding}
We therefore propose a compact circuit encoding for variational and non-variational gate-based quantum circuits.

The value \emph{and} type of each gate are left to an evolutionary algorithm. This means that rather than working with a fixed ansatz with variable parameters, the structure of the circuit is optimized concurrently with the parameters. 
The encoding of each gate $g_{i,l} \in [0;|\mathcal{GS}|), \; 0 \leq i < n \;, 0 \leq l < L$ (i.e., the gate on wire $i$ in layer $l$) is implemented as one individual mapping scalar, where we split the integer and the decimal parts of the floating-point value to realize the integer as the discrete choice of the gate-kind (mapped to a preselected, ordered gate-set $\mathcal{GS}$ of length $|\mathcal{GS}|$). We then use the decimal of the scalar to complete the selected gate according to one of three options:
\begin{itemize}
    \item[1] Gates with controls (e.g., \textit{CNOT}) at wire $i$ for gate $g_i$ place the control-target at the qubit corresponding to the normalized segment of the decimal, where each segment has an equal width $\frac{1}{n}$ and the range for segment $j \in [0; n-1)$ is defined as $ \left[\frac{j}{n}, \frac{j+1}{n}\right) \in [0;1], $ for all segments $[\text{segment}_0; \text{segment}_{n-1})$. In other words, the decimal selects the normalized segment (qubit), which corresponds to the normalized magnitude of the decimal $\in [0;1]$ between qubits $0$ to $n$. To safely include the boundary value 1, the last segment $i = n-1 $ is adjusted to $ \text{segment}_{(n-1)} = \left[\frac{n-1}{n}, 1\right]$. \\
    In the case that the corresponding target segment is the gate's wire itself, the action choice becomes the uncontrolled version of the gate (e.g., $\textit{CNOT}_{(i, j)} \to X_i$, $\iff i=j$).

    \item[2] Variational gates (e.g., $R_X(\beta)_i$) at wire $i$ for gate $g_i$ treat the decimal as the gate angle $\alpha$. To avoid  over-rotation, in the realization of the gates, we remap the decimal $[0;1] \to [0,2\pi]$.
    
    \item[3] Fixed operator gates (e.g., Hadamard $H$) without angles or targets discard the decimal, i.e., only the gate choice for gate $g_i$ remains.
\end{itemize}
The gate encoding of $n\times L$ scalars may then be flattened layer-by-layer (i.e., column-wise), to form our circuit encoding of dimensionality $n * L$, to be optimized by any search based optimization method. Since the only assumption made here is the preselection of a desirable gate set, this encoding is task agnostic and is generalizable to any ML/RL optimization or quantum control algorithm.

For the gate-selection we test the following gate-set $\mathcal{GS}$ combinations:
\begin{align*}
    \mathcal{GS}_\texttt{CliffordT} &:= \{\textit{CNOT}, H, S, T, I\}\\
    \mathcal{GS}_\texttt{RotCNOT} &:= \{R_X, R_Y, R_Z, \textit{CNOT}, I\}\\
    \mathcal{GS}_\texttt{TinyH} &:= \{R_X, H, \textit{CNOT}, I\}\\
    \mathcal{GS}_\texttt{Tiny} &:= \{R_X, \textit{CNOT}, I\}
\end{align*}
where \texttt{CliffordT} is the universal quantum gate set (cf. \citet{williams2010explorations}) and \texttt{RotCNOT},\;\texttt{Tiny}, \texttt{TinyH} are practically applied reductions, i.e., commonly used VQC building-blocks (cf. \citet{tilly2022variational}). All gate sets include the identity operator $I$ to allow for potentially sparse layer designs.

The quantum circuit is then constructed by mapping each element of the vector to a gate in the quantum circuit. Element $e_i$ of the vector maps to a gate at a fixed position within the circuit. In the $\mathcal{GS}_\texttt{Tiny}$ framework, a value of 3 or greater maps to an identity in the circuit at that position. If the value is $2\le e_i \le 3$, the inserted gate is an $R_X$ rotation gate with parameter $(e_i - 2)\cdot 2\pi$. In the last case, $1 \le e_i < 2$, a CNOT gate is inserted, with the target qubit number uniformly mapped to the fractional part. The other gate sets act similarly, with differences in the density of changes in the fractional parts.

\subsection{Quality Measures}\label{subsec:qd_measures}
Defining our QD measures, the objective function will be the combinatorial optimization hamiltonian cost objectives as described in the background section on CO. As our quality metrics, we choose \textit{circuit sparsity} and \textit{gate diversity} to search for circuits potentially resistant against barren plateaus for the following reasons: 

Over-parameterization of gates, in particular with the noisy NISQ era QC hardware, has not only been shown to make VQCs sensitive to barren plateaus \citep{wang2021noise} but also can -- depending on the ansatz design -- hinder the circuit expressivity wrt. state reachability in the Hilbert space \citep{larocca2024review}. As such, we want to explore the sparsity of our circuit solutions in the range $[0;n*L]$, where $0$ sparsity implies a fully parameterized qubit-to-layer mapping and $n*L$ is the empty circuit. Formally, the QD measure sparsity of solution $\theta$ is defined as the count of $$\text{sparsity}(\theta) = \sum_{l=0}^L \sum_{i=0}^n \mathbb{I}[\theta_{i, l} \neq I],$$ where $\mathbb{I}$ is the indicator function and $I$ is the Identity gate (i.e., an empty position).

Similarly, the concept of unitary symmetry, i.e., of the unitary spanned by the realized state of the circuit in the Hilbert space, is thought to relate closely to classical simulability \cite{cerezo2023does}, where completely symmetric unitaries offer no quantum effects, and, hence, are classically computable. Therefore a certain degree of non-uniformity in the circuit is desirable, which we express as the QD measure \textit{gate diversity} of solution $\theta$, or formally: $$ \text{gate diversity}(\theta) = \sum_{l=0}^L |\{ \theta_l \}| \setminus \{I\},$$ where $|\{ \theta_l \}|$ is the size of the unique element set of all gates on layer $l$ of solution $\theta$ (not counting Identity gates). This measure spans the range $[0; |\mathcal{GS}|*L]$, where a gate diversity of $0$ implies no gates in the circuit, up to every different available gate being set on every layer.

Our circuit samples are then optimized via QD-CMA search, using the respective CO Hamiltonian as objective function, and the sparsity and gate diversity as quality measures. We test our approach with different learning rates $\alpha$ to employ CMA-ES, CMA-ME or CMA-MAE, respectively.\footnote{Implementation of our approach can be found at \url{https://github.com/m-zorn/qd4vqc}}\\

\subsection{Baselines}\label{subsec:baselines}
Since there are numerous possibilities of iterating circuits, and testing all recent VQC proposals is out of scope for this paper (cf. \citet{sim2019expressibility} for a selection), we instead conform to the following two well-known, respectively state-of-the-art methods for both constructing and optimizing quantum circuits: For a gradient-based optimizer we choose the analytic Quantum Approximate Optimization Algorithm (QAOA) ansatz as well know benchmark algorithm \cite{farhi2014quantum}. For a gradient-free quantum optimization method, there are many recent examples, \cite{sunkel2023ga4qco, lukac2003evolutionary, ding2008evolving, chen2020variational}, however, we pick the quantum NEAT (QNEAT) \cite{giovagnoli2023qneat} algorithm as a baseline -- based on NEAT \cite{stanley2002evolving} for classical topological architecture search for their competitive performance to QAOA and comparable setup. For further details and formalization to both QAOA and QNEAT we refer the reader to their respective publications.

\section{Experiments}\label{sec:experiments}
\textbf{Gate-Set Choice} We begin our experimental evaluation with an ablation on the suitability of QD-optimization for the four different gate sets introduced previously, \texttt{CliffordT}, \texttt{RotCNOT}, \texttt{TinyH} and \texttt{Tiny}. We test the three CMA variants, CMA-MAE, CMA-ES, and CMA-ME, on solutions with $L=4$ circuit layers, $20$ emitters (batch size $5$ each). In this case study, we consider MaxCut problem graphs, with $8$ vertices, i.e., the solution circuits are of size $8\times 4 = 32$. The three test graph types (\textit{Barbell, Ladder, Caveman}) are considered in reference to \cite{khairy2020learning} as well as for direct comparability to \cite{giovagnoli2023qneat}. \textit{Barbell, Ladder} and \textit{Caveman} are characteristic graph instances in that they each represent distinct graph properties, i.e., complete graphs that are sparsely connected (Barbell, with subgraphs of size $p_B$), uniformly structured graphs (Ladder, with $p_L$ repetitions), and graphs consisting of cliques (Caveman, with $c$ cliques of size $|c|$, i.e., $p_C = (c, |c|)$). For gate-set suitability, we show the optimization quality via the solution approximation ratio on these three representative graph instances.

\begin{figure*}[t!]
    \centering
    \includegraphics[width=0.9\textwidth]{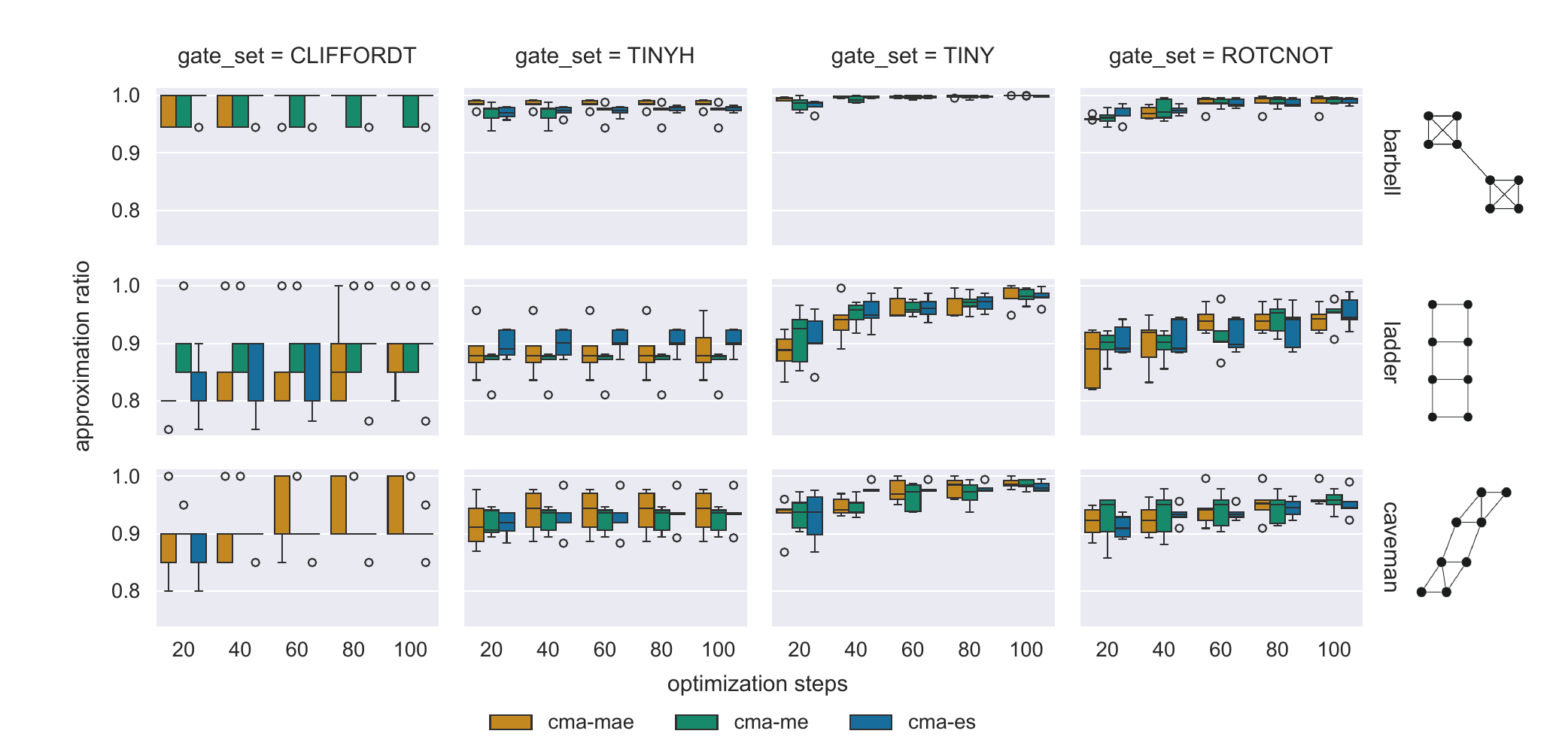}
    \caption{Ablation on the optimizability of different gate sets (from left to right: \texttt{CliffordT}, \texttt{TinyH}, \texttt{Tiny}, \texttt{RotCNOT} on the three example graphs barbell (top), ladder (center), caveman (bottom) after 100 optimization steps \textbf{(x-axis)}. The approximation ratio to the optimal objective energy is shown on the \textbf{(y-axis)}. Boxplot error bars show the 95\% confidence interval over $5$ runs each.}
    \label{fig:gate_ablation}
\end{figure*}

\begin{figure*}[ht!]
    \centering
    \includegraphics[width=0.7\textwidth]{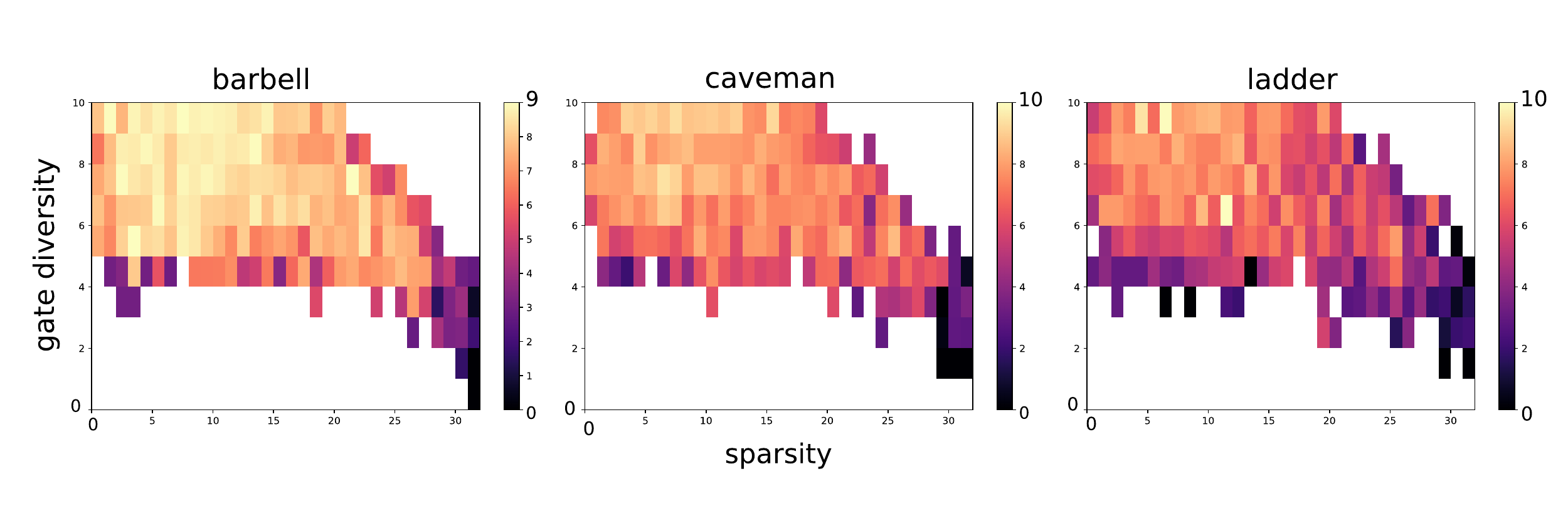}
    \caption{Single MaxCut run heatmaps of the elite archives after 100 steps of CMA-MAE optimization on the example graphs \textbf{(barbell, left)}, \textbf{(caveman, center)}, \textbf{(ladder, right)}. Lighter colors indicate higher objective function values for the quality criteria \textbf{(sparsity, x-axis)} and \textbf{(gate diversity, y-axis). Axes start with (0,0) bottom left.}
    }
    \label{fig:heatmaps}
\end{figure*}

The graphs and results are shown in Figure~\ref{fig:gate_ablation}. We observe that the variational gate sets \texttt{RotCNOT}, \texttt{TinyH} and \texttt{Tiny} significantly outperform the \texttt{CliffordT} set, with \texttt{Tiny} emerging as the variational gate set with the most variance, but also the best optimization potential. This result can be expected since \texttt{CliffordT} consists of mainly fixed gate operators, i.e., without parameterization, and is effectively only using half the scalar encoding compared to the parameterized control that variational gate sets allow. We also note that simply removing the Hadamard $H$ operator (from the \texttt{TinyH} set) seems to improve the CMA optimization process, indicating that small and basic operator sets like \texttt{Tiny} are favorable choices for the CMA methods.

 \begin{table}
    \centering
    \begin{tabular}{lcccc}
        \textbf{v=12} & maxCLI  & maxCUT & maxIND & minVER \\
        
         cma-es & 0.985 & 0.995 & 0.998 & 0.987 \\
         cma-mae & 0.973 & 0.993 & 0.996 & 0.976 \\
         cma-me & 0.984 & 0.997 & 0.998 & 0.979 \\\\

         \textbf{v=14} & maxCLI  & maxCUT & maxIND & minVER \\
         cma-es & 0.972 & 0.987 & 0.998 & 0.955 \\
         cma-mae & 0.965 & 0.980 & 0.995 & 0.942 \\
         cma-me & 0.982 & 0.979 & 0.996 & 0.956 \\\\
        
        \textbf{v=16} & maxCLI  & maxCUT & maxIND & minVER \\
         cma-es & 0.962 & 0.973 & 0.993 & 0.933 \\
         cma-mae & 0.961 & 0.968 & 0.992 & 0.900 \\
         cma-me & 0.964 & 0.976 & 0.991 & 0.912 \\

    \end{tabular}
    \caption{Scaling of different CMA-methods for graphs with $v=12,14,16$ vertices ($4$ layers). Each cell represents the average approximation ratio of $50$ randomly generated Erdos-Reny graph instances over $100$ optimization steps.}
    \label{tab:scaling}
\end{table}

\textbf{QD-measure Exploration} Considering three (singular but representative) optimization runs of the CMA-MAE method on the \texttt{TinyH} set in Figure~\ref{fig:heatmaps}, we can also observe which directions of the QD measures the CMA methods choose to explore and exploit. The heatmaps represent the result maps of the found elites (measure combinations), with color indicating their respective MaxCut objective score. In all three cases we find that CMA-MAE tends to explore towards sparser circuits with a high degree of gate variability. We conclude that the CMA, as such, naturally explores circuits less prone to barren plateaus (avoiding over-parameterization) but still leveraging quantum effects (with diverse layers, i.e., lowered symmetry). Heatmaps for CMA-ES and CMA-MA show similar trends, but not as pronounced as with CMA-MAE.

\begin{figure*}[ht!]
    \begin{subfigure}{\textwidth}
    \centering
        \includegraphics[width=0.85\linewidth]{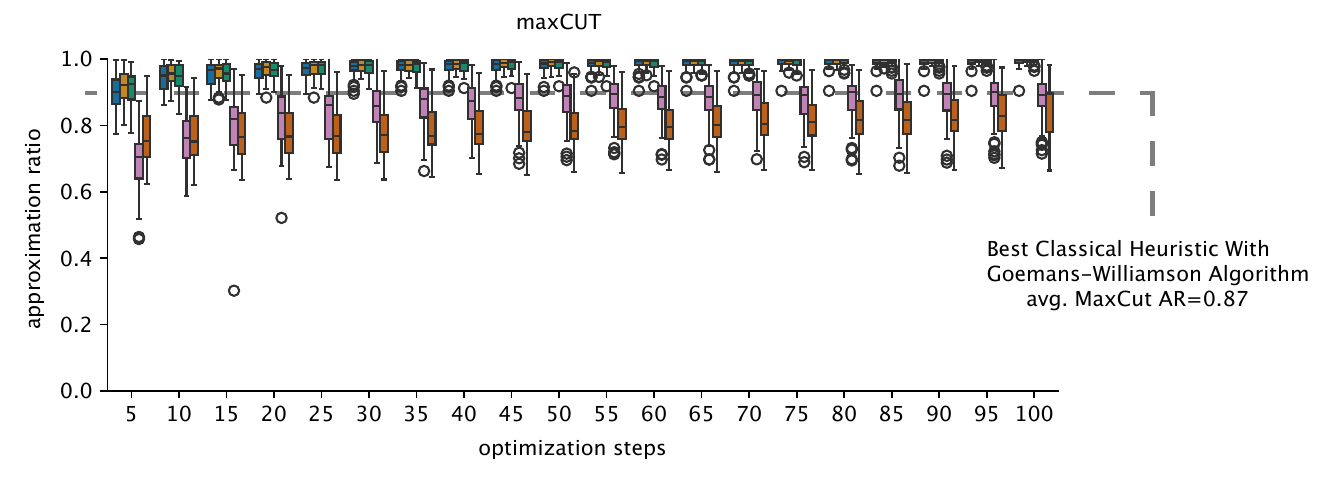}
        \caption{Evaluation MaxCUT problem}
        \label{fig:co_maxcut}
    \end{subfigure}
    \begin{subfigure}{\textwidth}
    \centering
        \includegraphics[width=0.75\linewidth]{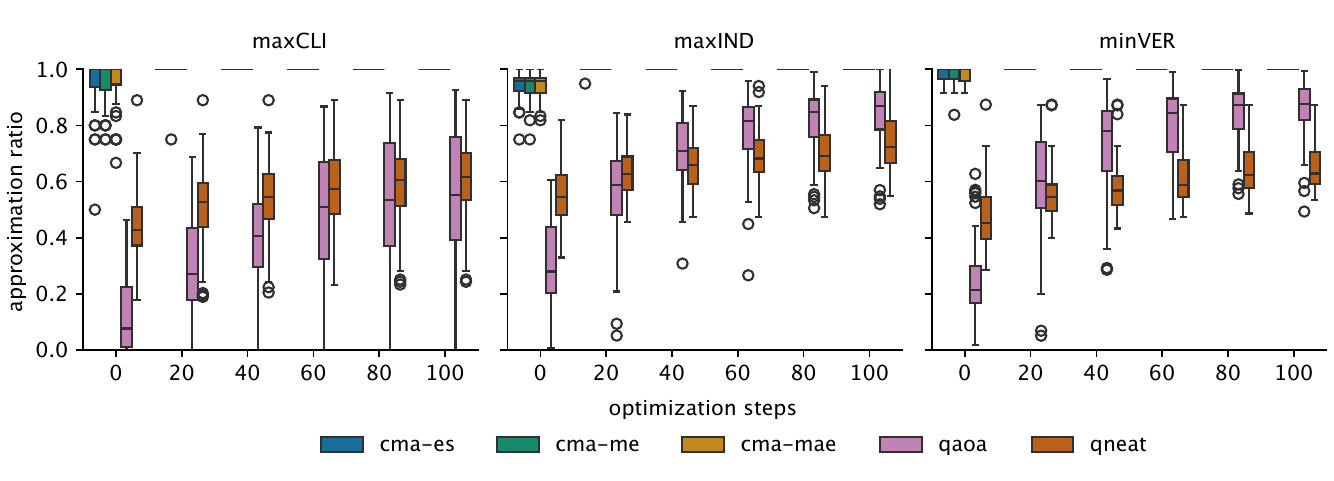}
        \caption{Evaluation MaxCLI, maxIND, minVer problems}
        \label{fig:co_other}
    \end{subfigure}
    \caption{
        Comparison of the optimization progress for the different combinatorial optimization problems, each on of 50 randomly generated erdos-reny graphs (connected), with 8 vertices (average edge-density $0.635 \pm 0.22$). Optimization steps over time are shown per graph, QNEAT with a population of $100$ individuals, QAOA with $p=4$, CMA-MAE (lr 0.3), CMA-ES (lr=0), CMA-ME (lr=1) producing 4 layer VQCs using the TINY gate set. Boxplots show optimization over $100$ steps, at every n-th step (x-axis), and the approximation ratios (y-axis) of the best solution energies found until this step / respective optimal solution energies. Error bars show the 95\% confidence interval.
        \textbf{(Top:)} Comparison of the optimization progress for the MaxCUT problem in greater detail (every $n=5$ optimization steps), \textbf{(Bottom:)} shows the MaxCLI, maxIND, minVer problems at every $n=20$-th step.
    }
    \label{fig:co_experiment}
\end{figure*}

\textbf{Baseline Evaluation} Finally, we conduct a direct baseline comparison to QAOA and QNEAT. We compare all three CMA variants (CMA-MAE, CMA-ES, and CMA-ME) to QAOA with $p=4$ layers, as well as QNEAT, with population sizes of $100$ individuals and the setup detailed in \cite{giovagnoli2023qneat}. We show the average approximation ratio (best-solution objective value / optimal-solution objective value) of $50$ randomly generated Erdos-Reny graph instances over $100$ optimization steps in Figure~\ref{fig:co_experiment}, first in detail for MaxCut in Figure~\ref{fig:co_maxcut} and as trend for the other three combinatorial optimization problems in Figure~\ref{fig:co_other}. We note that all CMA variants clearly and quickly outperform both baselines. All baselines do eventually find optimal solution circuits; however, QAOA only slowly, and QNEAT is characterized by very high variance. In comparison, CMA-MAE, CMA-ES, and CMA-ME perform relatively similarly, with CMA-MAE again slightly more optimal. We also note that QD methods are extremely more solution efficient, often finding optimal (or almost optimal) solutions within $20$ optimization steps. Efficiency on this scale can be considered quite valuable in both evaluation time and cost, considering that application on actual quantum hardware is still very expensive. A small scaling ablation of the CMA methods for problems graphs of $v=12,14,16$ vertices can be found in Table~\ref{tab:scaling} for comparison. While a slight degradation in approximation-ratio with increased vertex-size is notable, the general performance remains strong across the three CMA methods.

\subsection{Hardware Details}\label{subsec:appendix_hardware}
Experiments were simulated on two identical Linux (Debian 6.1) workstations, with AMD Threadripper 5995WX CPUs and NVIDIA 4090 GPU with 1024gb RAM.

\section{Limitations and Future Work}\label{sec:conclusion}
In this work, we have proposed and detailed a compact but expressive circuit encoding and shown that quality-diversity driven optimization with covariance matrix adaptation (CMA) methods notably outperforms current state-of-the-art optimizers a set of combinatorial optimization benchmark problems. We find that by appropriate design of our quality metrics (circuit sparsity and gate variance), QD methods will naturally explore and find circuit solutions that prevent the current NISQ era problems, such as barren plateaus and insufficient quantum effect through overly symmetric circuits. However, our work is still limited in scope, yielding many directions for future work. More extensive evaluation against QML-based methods on a variety of other benchmarks would round out this study with even more significance. Secondly, the choice of quality metrics could also be further explored, with concepts like \textit{circuit expressivity} and \textit{ciruit capacity}, which are also linked to barren plateaus \cite{larocca2024review}.

\section{Acknowledgments}
This paper was partially funded, by the German Federal Ministry for Economic Affairs and Climate Action through the funding program "Quantum Computing -- Applications for the industry" (contract number: 01MQ22008A), the German Federal Ministry of Education and Research through the funding program ``quantum technologies — from basic research to market'' (contract number: 13N16196) and through the Munich Quantum Valley, supported by the Bavarian state government's Hightech Agenda Bayern Plus.

\bibliography{bibliography}

\end{document}